# Toward Sensor and Software Product Line Based Context Aware Cloud Environment Assignment

**Asmae Benali[1], Bouchra El Asri[2] and Houda Kriouile[3]**

**IMS Team, SIME Laboratory ENSIAS, Mohamed V University, Rabat, Morocco**

## Abstract

Because of the growing interest for mobile device and pervasive applications deployed on cloud computing, the providing of intelligent and ubiquitous context-aware applications that take into account the user's context is one of the main challenges in future applications. In this article we consider how to augment applications aware context used by mobiles device and deployed on cloud computing. The behavior of these applications should depend not only on their internal state and user interactions but also on the context sensed during their execution. Indeed, our approach based on two essential mechanisms, context sensoring and context reasoning. We consider the information acquired by the context sensoring as a product line and we use feature models to represent this information received, the services provided by cloud provider, the available resources and constraints. At the context reasoning step, the context provisioning feature model (CSCAFM) algorithm permits to generate service context-aware that fulfils the requirements of use ensures a certain level of performance of resources.

***Keywords:*** *Cloud Computing, Context-Aware System, Context Sensoring, Context Reasoning, Software Product Line, Feature Model*

## 1. Introduction

Nowadays, a mobile device such as mobile phones has become the main sources of information for users these days. The number of mobile phone subscriptions is close to 4.6 billion [1] according the latest reports, which means that mobile phones are indispensable tools that we cannot live without. However, some applications deployed on mobile devices require more resources than a mobile device can offer. That's why, the connection of mobile device to an external source to obtain resources turn into necessary. One of such sources is cloud computing platforms which has recently emerged as a major trend in distributed computing where IT applications, platforms and infrastructures are provided as a service.

The use of context in mobile devices and ubiquitous computing research have recently received a rising attention. Effectively, Mobile users are increasingly demanding services that suit their context and the environment of the use. Many definitions of context notion in the literature have been provided. Dey et al. [2] they defined context as:

"... any information that can be used to characterize the situation of an entity. An entity is a person, place or object that is considered relevant to the interaction between a user and an application, including the user and applications themselves."

This has also been related to physical context, in order to differentiate it from the other types of context in mobile computing, like network infrastructure conditions [3].

Cloud computing should provide a context-aware services at different levels within a mobile device. At user level, the use of context facilitates the interaction between the human and the device. The second level concerns the application level where the context-awareness permits the adaptation of applications and enables context based services. At systems level, it can be used to manage the device resources such as the power, bandwidth, etc… and to improve the resource utilization.

Our model adapt a three-tier architecture, cloud users, broker and cloud service providers (CPSs). The broker is a server agent that manages the contextual knowledge of a pervasive computing environment [4]. It permits agents to access to a shared model of context and enables users to control the access of their information in a context-aware environment. There are two main essential mechanisms of context detection: context sensing which acquires information from the physical environment and context reasoning that interprets the information collected. The focus of this article is on how to provide intelligent, efficient and ubiquitous applications in mobile devices. In our approach, we consider the information acquired by the context sensing as a product line and we use feature models to represent these information in one hand. On the other hand, we represent services provided by cloud provider, the available resources and constraints using feature model (FM). Then, at the context reasoning step, the context provisioning feature model algorithm (CSCAFM) permits to generate service context-aware that fulfils the requirements of use and the minimum required quality of context.

The rest of this paper is organized as follows. Section 2 presents the background of our work. Section 3 describes the problem. In Section 4, we present our proposed solution. Section 5 illustrates our solution. In Section III we briefly compare and position our solution with other proposals found in the literature. Finally, in section VII, we conclude and introduce our future work.

## 2. Background

In this section, we define some of the concepts related to the present study.

### 2.1 Context-Aware Systems

Context-aware systems are a component of a ubiquitous computing or pervasive computing environment [5]. Dey and Abowd [2] defined Context-Aware as: "A system is context-aware if it uses context to provide relevant information and/or services to the user, where relevancy depends on the user's task". Context-aware systems are able to adjust their changing to a specific situation or condition without user intervention. In fact, they use context information to provide relevant services and information. Although location is a primary capability, Context-aware in contrast is used more generally to include nearby people, devices, lighting, noise level, network availability, and even the social situation.

### 2.2 Integration of sensors in Mobile Devices

Sensors have been an important addition to mobile devices, which can be enabled with their proper integration into all types of devices. This is usually aimed to facilitate capture of very specific context as input Multi sensor based context-Awareness [6].

### 2.3 Multi Sensor based Context-Awareness

By their nature, Context-aware devices need some sensors to react to the environment changes [7]. However, the consummation of resources and configuration are important requirements in mobile systems. While designing devices with sensors, price and performance often dictate sensor selection. Indeed, the management of resources and the selection of such configuration is a big challenge and should be properly addressed to enable context-aware platform.

### 2.4 Software Product Lines

The software product lines is a recent approach that favors systematic reuse throughout the software development process and enables the development of a set of software products with a considerable gain in terms of cost, time and quality. A software product line is a set of software-intensive systems sharing a common, managed set of features that satisfy the specific needs of a particular market segment or mission and that are developed from a common set of core assets in a prescribed way [8]. SPLE is based on two fundamental processes, domain engineering and application engineering alongside the management of variability and commonality. The domain engineering is the development of assets which will be used in the product line, whereas the application engineering is concerned with the construction of final products with specific requirements. The composition of core assets produced in domain engineering for software product lines enhances reuse and provides new models of software product lines. However, this task is far from easiness or obviousness, especially because these models incorporate variable elements. The following sub-section discusses the role of variability in the context of software product lines.

### 2.5 Variability in SPL

While browsing the literature about variability in software product lines, it may refer to essential and technical variability [9], external and internal variability [10], product line and software variability [11].

Feature-Oriented Software Development (FOSD) is a methodology for the design and construction of software product lines based on the separation of concerns [12], each concern is modularized in a separate component called feature. The commonalities and variabilities among products in the same domain can be expressed in terms of features. A feature is a prominent or distinctive and user visible aspect, quality, or characteristic of a software system or systems [13]. A feature is either [14] i) Mandatory, it exists in all products if the parent is selected; ii) Optional, it is not present in all products; iii) Alternative (One Of), it specializes more general feature; only one option can be chosen from a set of features iv) Or: One or more features may be included in the product.

### 2.6 Software Product Line and aspect-oriented programming

Aspect-oriented programming is a programming paradigm [15] that focuses on modularization of crosscutting concerns, grouped into the common denominator of 'Advanced Separation of Concerns'. It aims to add additional behavior to existing code without modifying the code itself (an advice), rather, it specifies separately which code is modified using a "pointcut" specification. Pointcut is an expression defined by an aspect to select the join points which specify how classes and aspects are related. When the program execution attains a join point which is selected by one or more pointcut expression. AOP implementations typically provide many advice types, like before, after, afterReturning, afterThrowing, and around. Nowadays many researches pointed out that AOP offers better modularity and changeability of SPL than other variability mechanisms [16]. In addition, AOP permits the improvement of evolvability and stability of variabilities in an SPL at dynamic scenarios [17]. Aspects can help to modularize variabilities and make their addition or removal easier during the configuration of the application. We exploit this relationship between SPL and AOP in our approach to cope with the adaptation of the application at runtime like cloud services [18].

## 2.7 Cloud-based Context Services

In cloud environments, context consumers are context data considered like inputs to adjust their behavior to the users' current status (e.g. context aware activity). Regarding Context provider is an entity which affords context information by collecting raw data from context sources (e.g., captures, networks, mobile phones, etc) and convert them to significant and important information, aggregation, modeling, reasoning and download it to the context broker. The latter plays an intermediary role to control context flow between Context providers and Context consumers. Its principal function is the registration of available Context providers, to examine resolution, event management, afford routing and look for services [19][20]. The utilization of context information in cloud environment is devised in two cases [21]. The first context service is made to direct and afford improved service to the user. According the second case, the role of context is to detect faults and maintain the stability of system by adapting the operations that suit the situations of environmental change.

## 2.8 Quality of Service

Quality of Service (QoS) [22] presents an interesting topic in cloud computing which is a defined measure of performance in a system. It usually used to control the resource reservation so as to assure the availability of services and also to ensure certain level of performance. For instance, to make sure that real-time video is delivered without annoying blips, a traffic contract is negotiated between the consumer and cloud provider that guarantees a minimum bandwidth along with the maximum delay in milliseconds that can be tolerated. QoS models are associated with consumers and providers through the use of Service Level Agreements (SLA).

## 2.9 Quality of Context

Quality-of-Context (QoC) is "any information that describes the quality of information that is used as context information. Thus, QoC refers to information and neither to the process nor the hardware component that possibly provide the information"[23]. The most important QoC parameters outlined are the follwing: Precision, Probability of Correctness, Trust-worthiness, Resolution, Up-to-dateness [23, 24].

# 3. Problem Description

Recently cloud computing has an increasing in number of mobile devices that deployed, that's why the demand of context-aware services to attribute increases. Mobile Device should make information and services available taking into account the change of situations and the constraints of both the consumer and the service. Indeed, the context of mobile devices application is highly dynamic allowing the application to tackle with the constraints of mobile devices like presentation, communication restrictions and interaction abilities. Context-aware computing promises to capture the user's needs and hence the requirements they have on systems. However, several mobile devices suffer from a miss of computing resources such as memory, computing power, etc...Moreover, cloud computing has new challenges to confront when demanding information about user's environment in terms of location-aware, device-aware, time-aware, and personalized applications to meet with the constraints of mobile devices in matters of interaction abilities, resources and communication restrictions. In addition, the consumer also needs to check some information about the services proposed such as the availability of service, cost, response time of service, quality of the service and quality of context information. Then, the provider has to deal with this cloud variability and resources dimensions, e.g., database size.

# 4. Outline of the Approach

## 4.1 Architecture

In cloud third-tiers architecture, brokers provides the intermediary between cloud providers and cloud consumer that assist companies in choosing the services and offerings that best suits their needs [4]. They may also assist in the deployment and integration of applications across multiple clouds. The internal architecture of the context provisioning system on a cloud infrastructure is presented in figure 1.

## 4.2 Process

The process consists of four main activities that must take place so as to apply the approach (Figure 2). The output of these activities is a configuration file defining the feature model that represents the aware service that suits the context conditions, user's requirements, quality required, etc…

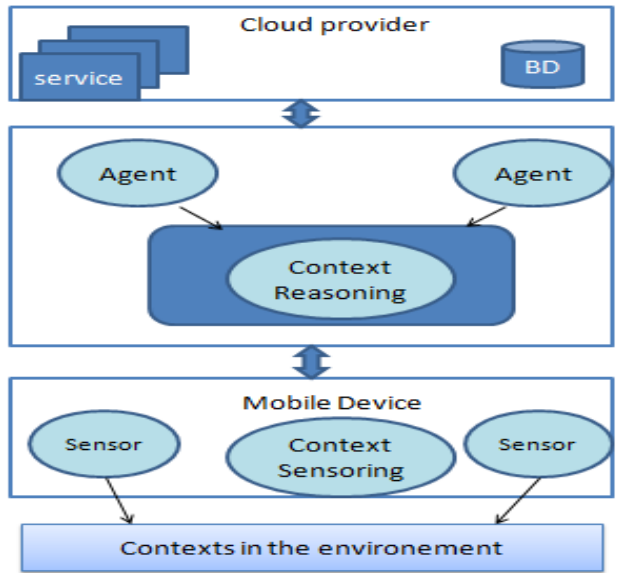

Fig. 1 Overall Architecture of the cloud based context-aware information

- Acquiring consumer context information: during this activity, sensors acquire the consumer information from the physical environment which is necessary to fix and contribute in the selection of the appropriate service needed.

-  Presenting the information acquired as FM: Once the information is acquired by the sensors, these latter will be modeled as a feature model where every context information is presented as a feature and the sub-information as a child features.

- Modeling the appropriate services as FM: When the consumer needs a service from the cloud provider, it sends a request to the broker entity which chooses the convenient services that might fulfil the requirements of the demand with their resources diminutions. Then, it presents these selected services as a feature model.

- Browsing and interpreting the FM constructed: In the context reasoning, the broker browses and interprets, by calling the CSCFM algorithm, the feature models acquired from the consumer and the provider entity. Then, it generates the context-aware service that answers to the requirements of the consumer and minimize the required quality of service. Furthermore, it controls the resource reservation to ensure the availability of the service and certain level of performance of resources like power, storage and other aspects such as reliability, security, etc…

### 4.2.1 Context Broker

A Broker Context is an intermediate entity that permits consumers to express their concern in order to obtain context information about services provided and it allows the provider to acquire information context about consumers (Figure 3). To model this variability in context information and build the efficient configuration that meet with the requirements of the consumers and the services providers, context broker bases on feature model mechanism. The context provisioning feature model (CSCAFM) algorithm permits to generate service context aware that fulfils the minimum required quality of context. Algorithm 1(Figures 4, 5 and 6) initializes the CSCAFM generation algorithm and returns the final result. It creates the root context specific feature and, for each sub-feature of the feature model, calls the procedure

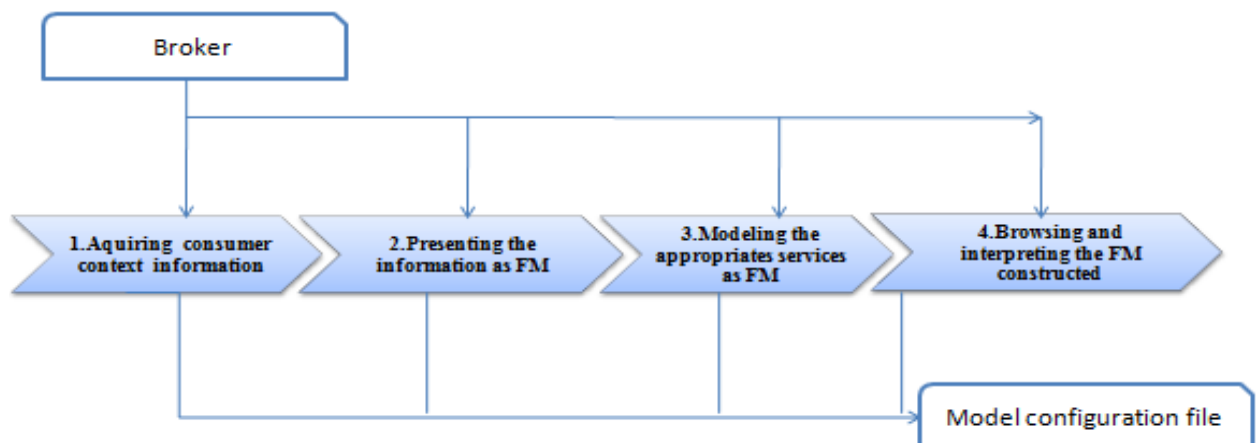

Fig. 2 Process of our proposed solution

DetectConsumerContext and SelectServiceToAdapt.

### 4.2.2 Consumer

In our approach, the consumer produces dynamic results according to the 5 WH questions: who, where, when, what, and why it was invoked. The context information of the consumer is presented as a service oriented product line using feature model to react to the changing condition of the context. It can react in many situational circumstances such as:

```
Procedure: DetectConsumerContext ( f_C )
Input: A feature  f_C  from that the consumer is
traversed
Result: Detect information consumer context change
foreach child feature f_child of  f_C  do
  DetectConsumerContext ( f_C )
end
```

Fig. 5 DetectConsumerContext algorithm

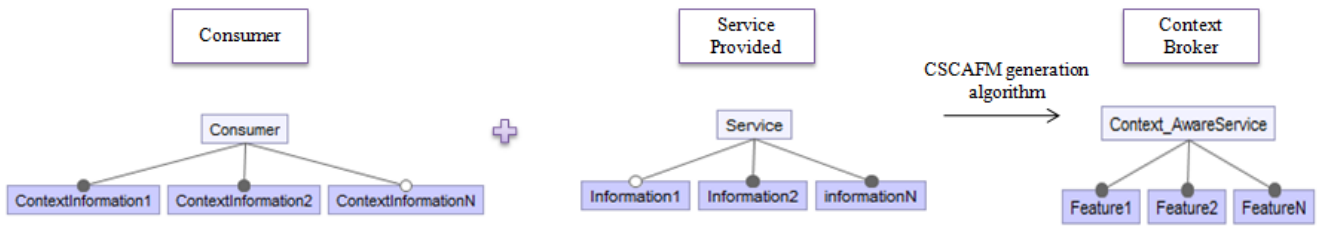

Fig. 3 Overview of context-aware service generation

- The device used by the client to invoke the service.
- The location of the client.
- The time at which the client invokes the service.
- The activity that the client is carrying out at the time it invokes the service.
- The preferences that the client may has defined before the service invocation.
- The security associated with client of the service.

```
Algorithm 1: Main procedure of the CSCAFM
generation algorithm.
Input: the context and feature models
Result: A CSCAFM model built according to the
context and feature models
Initialize an empty CSCAFM.
DetectConsumerContext ( f_C )
foreach child feature  f_s  of the FM root feature of
service  r_sf  do
  SelectFeatureServiceToAdapt( f_s , r_CPFM )
end
end
```

Fig. 4 CSCAFM algorithm

```
Procedure: SelectFeatureServiceToAdapt ( f_s , r_CSCAFM )
Input: A feature  f_s  from which  the service is
traversed
Result: select service and adapted it to the context of
the user
foreach child feature f_child of  f_s  do
if feature has child features context information then
  if  feature  has  mandatory  child  features  context
  information then
  foreach  child_feature  context  information  is  a
  mandatory  child  feature  of_feature  do
  _feature.QoC ← feature.QoC + CalculateMin_QoC_req
  uired (child_feature_context_information)
  SelectFeatureServiceToAdapt( f_s , r_CSCAFM )
  end
  end
end
```

Fig. 6 SelectFeatureServiceToAdapt algorithm

### 4.2.3 Service

The Service entity, in our framework, shows interfaces to the consumer via Broker entity in order to opt to context information and allows also the consumer to negotiate SLAs. The services are modeled by using feature model where each service is identified by its response time, cost, the computation resources required to run the service, quality of service and quality of context. As SaaS model is deployed over the internet and delivered to thousands of users, this model is the most appropriate to handle context provisioning in cloud computing.

### 4.2.4 Interaction Model

In this section, we present the interactions among the components of the framework. Figure 7 describes the interactions among these components. In fact, when the Consumer asks a service to the Broker context, this later tries to find the demanded service from the Service component and then demands some needed information to the Consumer service concerning the behavior context of the consumer. Afterward, the Broker context exposes interfaces to the Consumer entity to choose the context information that suits to the situational conditions. Then, the Broker context component generates the context-aware service, registers the generated service to keep the track of the operation and finally sends it to the Consumer component.

At first, when a consumer needs a service, the Consumer component invokes the getContext-awareService() method of the Broker context that forwards the request to Services component that are offering the service that fits the consumer's requirements. The Broker context receives notifications on quality of service offering change calling its notify() method that a Service component invokes. Afterward, it notifies a Consumer component about that change by invoking its notify() method. Then, the Broker context entity demands some needed information to the Consumer of this service concerning the behavior context of the consumer. The Broker context receives notifications on quality of context offering change using its notify() method that a Consumer component invokes. Afterward, it notifies a Consumer component about that change by invoking its notify() method. After that, the Broker context component invokes the findServiceContext () method of the Service component to get context information about the service. The Broker context receives notifications on context information calling its notify() method that a Service component invokes. The Broker context has also two additional methods generateContext-aware() and enregistrateContext-aware(). Regarding to the first method is invoked to generate the context-aware service. The second method is called to register the generated service in order to keep the track of the operation. Finally, the Consumer component invokes the sendContext-awareService() method of the Broker context to get the context aware service demand that its interest to use.

## 5. Case Study

In this work, we have adapted the SPL-based development in order to support specificities of both the cloud based application.

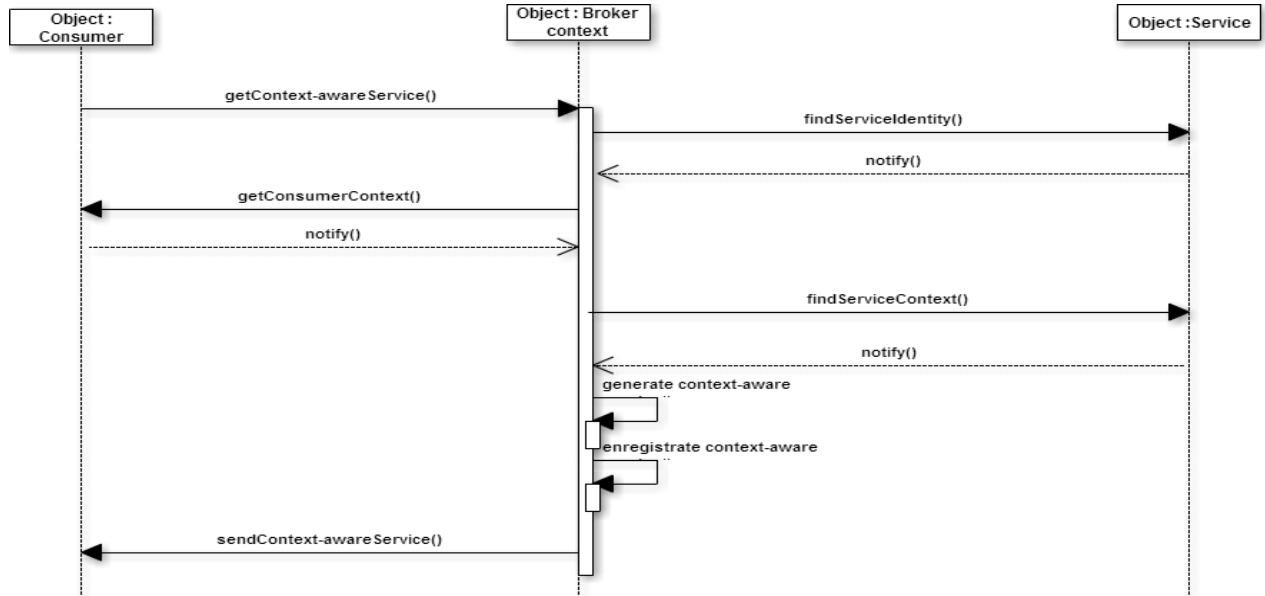

Fig. 7 Diagram of interactions among the framework actors

and the user's requirements using an extend feature model to introduce attributes to the features, where an attribute is a characteristic of a feature which can be measured. We have also introduced the notion of properties, which have the form of triples <name, type, value> that affected to a feature. The property can represent the cloud services information like pricing, response time, resources, QoS, QoC. Similarly, the property can represent the consumer information such as location, security, activity, mobile device. In this case study, we use the FeatureIDE tool [25] are alternative feature groups for Geolocation feature. We consider this running example, where the consumer sends a request to the broker entity in order to demand a service with the following requirements model: {< MachineSeize, Small >, < Authentication, OAuth >, <Geolocation, Europe>}. The broker gets from the cloud provider the services that fit the demand. Let us now consider the following service feature model (Figure 9). This model contains mandatory features such as, Cost, ResponseTime, Resource, QoS. It contains one optional feature, QoC. The to model our case framework for software product line engineering, the principal benefit of this tool is the possibility to generate the XML scheme of the feature model. In order to illustrate such approach, we have developed CC-SPL (Consumer Context Software Product Line) and CSC-SPL (Cloud Service Context Software Product Line). Figure 8 presents the Context information of a consumer (CC_SPL) extended feature model. This model contains mandatory features, like Id, Security, Mobile device, ProfileLanguage and Geolocation. There broker gets from context sensoring the information context about the consumer, the Geolocation which is Europe in this case to determine the cost, the unity of the cost, here Euro/hour and the value of the RAM, here SmallRam that should be affected to the consumer. Also it specifies related dimension's value when required. We have adopted the conditional compiling technique to enable variation in the proposed SPL. In conditional compiling, preprocessor directives designate the lines of code which should be compiled or not depending on the value of preprocessor variables. This decision can be at level of a line or a set of lines of code (Figure 10) or to a complete source code file. Because our running example was implemented using the Java programming language and this later has not native support for preprocessor directives, we have integrated an Eclipse IDE plug-in called Antenna Preprocessor [26] to annotate the source code. On the other hand, we aim to provide a flexible and dynamic solution that enables the variabilities to be woven and unwoven at the runtime.

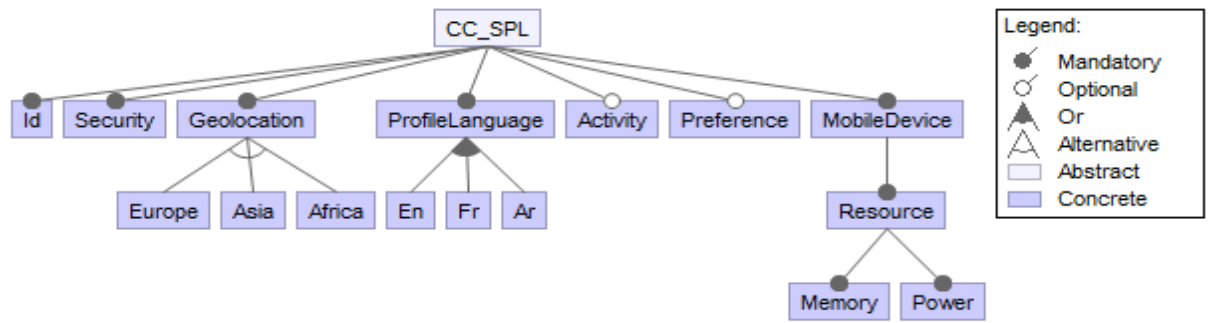

Fig. 8  CC_SPL feature model

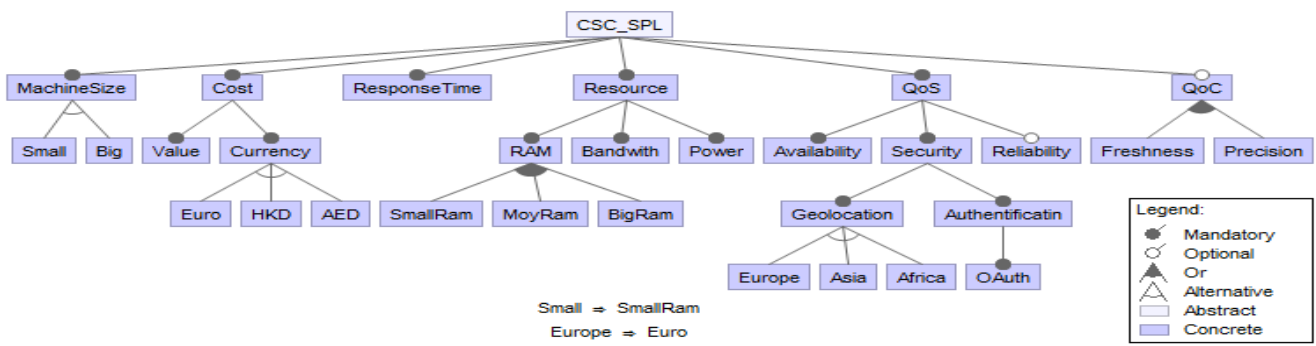

Fig. 9 CSC_SPL feature model

```
1  public class MachineSeize {
2      IMachineSeize returnValue = null ;
3
4  if(Price.isCalled())
5⊖ {
6
7      try{
8          //Small Machine feature
9          //#if Small
10 //@      returnValue=MachineSmall.getPrice();
11         //#endif
12         //Big Machine feature
13         //#if Big
14 //@      returnValue= MachineBig.getPrice();
15         //#endif
16         returnValue.adapte();
17         }
18     catch(SecurityMechanismException e){
19         e.printStackTrace();}
20
21     }
22  }
```

Fig. 10 Example of variability with conditional compiling in the Security

For these reasons, we use JBoss AOP framework [27] for implementing our solution, which supports static and runtime weaving. This framework permits to insert aspects for static changes via XML annotations or annotations and regarding the runtime changes, it uses the JBoss AOP API. JBoss AOP bases on code instrumentation in order to support the runtime binding and unbinding of aspects. Figure 11 presents a fragment of the XML representation of the CSC_SPL feature model that uses the Geolocation variability that associates to the Security feature. The lines 3 to 13 show the pointcut, aspect and advice for the Geolocation variability.

```
1  <product>
2     <feature name="Geolocation" >
3        <pointcuts>
4           <pointcut name= "GeoPointcut">
5              execution(lib.security.SecurityMechanism-> SecurityApply())
6           </pointcut>
7        </pointcuts>
8        <bindings>
9           <before pointcut="GeoPointcut">
10             aspect="CC_SPL.aspects.security.UserSecurity"
11             name="VerifyUserSecurity"
12             />
13          </bindings>
14       </feature>
```

Fig. 11 XML representation of the product description

## 6. Related work

This section presents several researches for managing context-aware systems with a distributed computing. Sensor-based systems sense and incorporate data about people, time, noise level and location, as well as many other pieces of information to adjust their model of the user's environment [28, 29]. The Active Badge system which integrates the position technique with a distributed computing infrastructure [30]. The ParcTab experiment is investigation into context-aware computing where the mobiles devices were augmented with locality for mobile access to location-based services [31]. However, all these works have used indirect awareness where sensors located in the infrastructure and engaged to listen for beacons sent from the mobile devices.  Several other context-aware mobile systems have employed cell-of-origin like context to locate the services, such as the GUIDE system deployed on the basis of a wireless LAN [32]. Traditional methods relied on cross-modality error correction that does not fit to pervasive computing applications [33]. [34] proposes context-based techniques which do not demand the interventions of the user. They are many approaches that have used context to represent human reactions, mostly in agent-based systems [35, 36]. These approaches are based on the assumption where the expectations are implicit when a person or an agent is in a known context. Capucine [37] presents a Model-Driven Engineering (MDE) approach for using SPLs to expose the context information. But, these works cope context-awareness with the MDE approach and then are used especially for transformation purposes.

## 7. Conclusions

In this paper, we describe a novel approach to context modeling for cloud environment scheduling services. It is based largely on Dynamic Software Product Line and context-aware information, showing how it can be interfaced with sensors by sensing context information of consumer and its environment. Indeed, the context behavior change of both the consumer and the services generated by cloud provider. Sensors acquire information about consumer and device mobile environment; this information will be modeled as a FM. To response to consumer's request, the broker browses and interprets, by calling the CSCAFM algorithm, the FMs described, adapted and reconfigured in order to generate the context-aware service that suits the requirements of both the consumer and cloud provider. It also optimizes, controls the resources reservation and assures a certain level of performance of resources. In ongoing work, we plan to propose a mechanism that verifies the correctness and consistency of feature models for mobile and context–aware SPLs based on OCL.

## References

[1]  http://en.wikipedia.org/wiki/Mobile_Phones.Wikipedia,"Mo bile phone - Wikipedia, the free encyclopedia," Accessed 19 Mars 2016.

[2]  A.Dey, and G. Abowd, "Towards a better understanding of context and context-awareness", in CHI 2000 Workshop on The What, Who, Where,When, and How of Context-Awareness, 2000.

[3]  A. Sheth, and C. Henson, S.S. Sahoo, "Semantic sensor web. IEEE Internet Comput", 2008, 12, 78–83.

[4]  R. Buyya, C. Shin Yeo, S. Venugopal, J. Broberg, and I. Brandic, "Cloud computing and emerging {IT} platforms: Vision, hype, and reality for delivering computing as the 5th utility", Future Generation Computer Systems, 25(6):599 – 616, 2009.

[5]  R. Hull, P. Neaves, and J. Bedford-Roberts, "Towards Situated Computing; International Symposium on Wearable Computers", 1997, pp.146-153.

[6]  J. Healey,and R. Picard, "StartleCam: A Cybernetic Wearable Camera", Proc. of the International Symposium on Wearable Computing, Pittsburgh, PA, USA, October 1998, pp. 42-49.

[7]  A. Schmidt, and K. Van Laerhoven, "How to Build Smart Appliances?", IEEE Personal Communications 8(4), August 2001.

[8]  P. Clements, and L. Northrop, "Software Product Lines: Practices and Patterns", Addison-Wesley Longman Publishing Co., Inc., 2001.

[9]  G. Halmans, and K. Pohl, "Communicating the variability of a software product family to consumers", Software and Systems Modeling, 2(1):15–36, 2002.

[10] K. Pohl, G. Böckle, andF. J. van der Linden, "Software Product Line Engineering: Foundations, Principles and Techniques", Springer-Verlag New York.

[11] A. Metzger, K. Pohl, P. Heymans, P. Schobbens, and G. Saval, "Disambiguating the Documentation of Variability in Software Product Lines: A Separation of Concerns, Formalization and Automated Analysis",

[12] D.L. Parnas, "On the criteria to be used In decomposing systems into modules", in Software pioneers, B. Manfred

[13] K.C. Kang, S.G. Cohen, J.A. Hess, W.E. Novak, and A.S. Peterson,"Feature-oriented domain analysis (FODA) feasibility study", DTIC Document,1990.

[14] J.V. Gurp, J. Bosch, and M. Svahnberg, "On the notion of variability in software product lines" in Proceedings of the Working IEEE/IFIP Conference on Software Architecture, pp. 45, 2001.

[15] R. Filman et al., "Aspect-Oriented Software Development", USA, Addison-Wesley, 2005.

[16] E. Figueiredo et al., "Evolving software product lines with aspects: An empirical study on design stability", Proc. of the 30th Int. Conf. on Software Engineering. USA: ACM, 2008, pp. 261-270.

[17] S. Apel and D. Batory, "When to use features and aspects? A case study", Proc. of the 5th Int. Conf. on Generative Programming and Component Engineering. USA: ACM, 2006, pp. 59-68.

[18] T. Dinkelaker et al., "A dynamic software product line approach using aspect models at runtime", Proc. of the First Workshop on Composition and Variability. USA: ACM, 2010.

[19] H. Vahdat-Nejad, K. Zamanifar, and N. Nematbakhsh, "Towards a Better Understanding of Context Aware Middleware: Survey and State of the Art", To be published, 2013.

[20] A.S.L. Kiani, A. M. Knappmeyer, N.Bessis, and N. Antonopoulos, "Federated broker system for pervasive context provisioning", Journal of Systems and Software, vol. 86, pp. 1107-1123, 2013.

[21] H. Jung, and S. Dong, "A Conceptual Framework for Provisioning Context-aware Mobile Cloud Services", Cloud Computing (CLOUD), 2010 IEEE 3rd International Conference on, 2010.

[22] D. Armstrong, and K. Djemame, "Towards Quality of Service in the Cloud", School of Computing, University of Leeds, United Kingdom.

[23] T. Buchholz, A. Kpper, and M. Schiffers, "Quality of context: What it is and why we need it?", In Proc. of the 10th International Workshop of the HP OpenView University association (HPOVUA), 2003.

[24] K. Sheikh, M. Wegdam, and M. Van Sinderen, "Quality-of-Context and its use for Protecting Privacy in Context Aware Systems", Journal of Software, vol. 3(3) pp. 83-93, March 2008.

[25] C. Kastner, T. Thum, G. Saake, J. FeigenspanLeich, T.F. Wielgorz, and S. Apel, , "Featureide: A tool framework for feature-oriented software development", In IEEE 31st International Conference on Software Engineering, IEEE 2009, pp. 611–614.

pioneers, B. Manfred In 15th IEEE International Requirements Engineering Conference, RE '07, pages 243–253, Oct 2007.

[26] Antenna Preprocessor:
http://antenna.sourceforge.net/wtkpreprocess.php, Accessed 1 August 2016.

[27] JBoss AOP Framework: http://www.jboss.org/jbossaop, Accessed 2 August 2016.

[28] G.D. Abowd, and E.D. Mynatt, "Charting past, present, and future research in ubiquitous computing", ACM Trans. Comput. Human Interact. 2000, 7, 29–58.

[29] A. Oikonomopoulos, I. Patras, and M. Pantic, "Human action recognition with spatiotemporal salient points", IEEE Trans. Syst. Man Cybernet. 2006, 36, 710–719.

[30] R. Want, A. Hopper, J. Falcao, and V. Gibbons, "The Active Badge Location System", ACM TIS, 1992, pp. 91-102.

[31] R. Want, B.N. Schilit, I.N. Adams, R. Gold, K. Petersen, D. Goldberg, R.J. Ellis, and M. Weiser, "ParcTab Ubiquitous Computing Experiment", Technical Report CSL-95-1, Xerox Palo Alto Research Center, March 1995.

[32] N. Davies, K. Cheverst, K. Mitchell, and A. Friday," Caches in the Air: Disseminating Information in the Guide System", Proc. of the 2nd IEEE Workshop on Mobile Computing Systems and Applications (WMCSA'99), New Orleans, Louisiana, USA, 1999.

[33] R. Khosla, I. Sethi, and E. Damiani, "Intelligent Multimedia Multi-Agent Systems: A Human-Centered Approach", Kluwer, MA, USA, 2000.

[34] Y. Sakurai, S. Hashida, R. Tsuruta, and H.S. Ihara, "Reliable Web-Based CSCW Systems Using Information Fusion of Various Multiple Biological Sensors", In Proceedings of the 4th International IEEE Conference on Signal Image Technology and Internet Based Systems (SITIS' 2008), Bali, Indonesia, 30 November–3 December 2008, pp. 480–489.

[35] V. Akman, and M. Surav, "Steps toward formalizing context", AI Mag. 1996, 17, pp. 55–72.

[36] A.J. Gonzalez, and R. Ahlers, "Context-based representation of intelligent behavior in training simulations", Trans. Soc. Comput. Simulat. 1996, 15, pp.153–166.

[37] C.Parra, X. Blanc, and L. Duchien, "Context Awareness for Dynamic Service-Oriented Product Lines", In: 13th International Software Product Line Conference (SPLC), San Francisco, CA, USA , 2009.

**Asmae Benali** received engineer degree in computer science and software engineering in 2013 from the National Higher School for Computer Science and Systems Analysis (ENSIAS) at Mohammed V University of Rabat, Morocco. She is actually working on Ph.D at ENSIAS School. Her main researches are Cloud Computing, Context-aware Systems, Autonomic Systems, Software Product Line Engineering, Service-Oriented Computing.

**Bouchra El Asri** is a Professor in the Software Engineering Department and a member of the IMS (Models and Systems Engineering) Team of SIME Laboratory at National High School of Computer Science and Systems Analysis (ENSIAS), Rabat. Her research interests include Service-Oriented Computing, Model-Driven Engineering, Cloud Computing, Component-Based Systems and Software Product Line Engineering.

**Houda Kriouile** received her engineer degree in computer science and software engineering in 2012 from the National Higher School for Computer Science and Systems Analysis (ENSIAS) at Mohammed V University of Rabat, Morocco. She is a Ph.D. candidate, and member of the IMS team from the SIME laboratory of ENSIAS School. Her research interests include Component-Based Systems and Software Product Line Engineering, Cloud Computing, Service Component Architecture.